\documentclass[epj]{svjour}
\usepackage{graphics}
\begin{document}
\title{Prospects for Discovering Supersymmetry at the LHC}
\author{John Ellis\inst{1} 
}                     
%
%
\institute{Theory Division, Physics Department, CERN, CH-1211 Geneva 23, Switzerland} 
%
\date{Received: date / Revised version: date}
%
\abstract{
Supersymmetry is one of the best-motivated candidates for physics beyond the
Standard Model that might be discovered at the LHC. There are many reasons to
expect that it may appear at the TeV scale, in particular because it provides a natural
cold dark matter candidate. The apparent discrepancy between the experimental
measurement of $g_\mu - 2$ and the Standard model value calculated using low-energy
$e^+ e^-$ data favours relatively light sparticles accessible to the LHC. A global
likelihood analysis including this, other electroweak precision observables and B-decay
observables suggests that the LHC might be able to discover supersymmetry with
1/fb or less of integrated luminosity. The LHC should be able to discover
supersymmetry via the classic missing-energy signature, or in alternative phenomenological
scenarios. The prospects for discovering supersymmetry at the LHC look very good.
\PACS{
      {12.60.Jv}{Supersymmetric models}
         \and
      {14.80.Ly}{Supersymmetric partners of known particles}
     } 
} 
\maketitle
\begin{center}
CERN-PH-TH/2008-208
\end{center}

\section{Introduction}
\label{intro}

Many theorist-years of effort have been invested in work to understand
the structure of supersymmetry and how to construct realistic supersymmetric models,
as well as how to test them experimentally. Many experimentalist-years have also
been invested already in the simulation of possible supersymmetric signals at
the LHC, and many more will be invested in searches once the LHC starts to
provide high-energy collisions. What are the prospects that all these efforts will be
crowned with success? What follows is my personal perspective on this question,
with illustrations taken largely from papers in which I have been a co-author: I
apologize in advance for not referring correctly to all the relevant literature.

\section{Motivations}
\label{Mots}

One's assessment of the LHC's prospects for finding supersymmetry
depends to great extent on the reasons why one
might think that supersymmetry exists, in particular at accessible energies.
There are many idealistic motivations for believing in supersymmetry, such as
its intrinsic elegance, its ability to link matter particles and force carriers, its
ability to link gravity to the other fundamental interactions, its essential role
in string theory, etc. However, none of these aesthetic motivations gives any
hint as to the energy scale at which supersymmetry might appear.
There are, however, various utilitarian reasons to think that supersymmetry
might appear at some energy accessible to the LHC. 

Historically, the first of
these was the observation that supersymmetry could help stabilize the mass
scale of electroweak symmetry breaking, by cancelling the quadratic
divergences in the radiative corrections to the mass-squared of the Higgs boson~\cite{hierarchy},
and by extension to the masses of other Standard Model particles. This
motivation suggests that sparticles weigh less than about 1~TeV, but the exact 
mass scale depends on the amount of fine-tuning that one is prepared to
tolerate. 

The second motivation for low-scale supersymmetry
was the observation that the lightest supersymmetric particle (LSP) in
models with conserved $R$ parity, being heavy and
naturally neutral and stable, would be an excellent candidate for the dark
matter that clutters up the Universe~\cite{EHNOS}. This motivation requires that the lightest
supersymmetric particle should weigh less than about 1~TeV, if it had once
been in thermal equilibrium in the early Universe. This would have been the
case for a neutralino $\chi$ or a sneutrino $\tilde \nu$ LSP, and the argument can
be extended to a gravitino LSP because it may be produced in the decays of
heavier, equilibriated sparticles.

The third reason for thinking that supersymmetry may appear within the LHC
energy range is the observation that including sparticles in the 
renormalization-group equations (RGEs) for the gauge couplings of the Standard Model
would permit them to unify~\cite{GUT}, whereas unification would not occur if only the
Standard Model particles were included in the RGEs. However, this argument does not
constrain the supersymmetric mass scale very precisely: scales up to about
10~TeV or perhaps more could be compatible with grand unification.

The fourth motivation is the fact that the Higgs boson is (presumably)
relatively light, according to the precision electroweak data. It has been known
for some 20 years that the lightest supersymmetric Higgs boson should weigh
no more than about 140~GeV, at least in simple models~\cite{EGZ}. For around 15 years now,
the precision electroweak noose has been tightening, and the best indication now
(incorporating the negative results of searches at LEP and the Tevatron) is that
the Higgs boson probably weighs less than about 140~GeV~\cite{EWWG}, in perfect agreement with
the supersymmetric prediction.

Fifthly, if the Higgs boson is indeed so light, the present electroweak vacuum would be
destabilized by radiative corrections due to the top quark, unless the Standard Model is 
supplemented by additional scalar particles~\cite{ER}. This would be automatic in supersymmetry,
and one can extend the argument to `prove' that any mechanism to stabilize the electroweak 
vacuum must look very much like supersymmetry.

For all these reasons, I believe there is a high likelihood that supersymmetry will appear
at or near the electroweak scale. The next question is whether, even accepting all these
arguments, it is nevertheless guaranteed to appear at the LHC.

The clinching argument could be provided by the ano- malous magnetic moment of the
muon, $g_\mu - 2$. As is well known, the experimental measurement of this quantity~\cite{g-2} disagrees with the Standard Model prediction~\cite{g-2th}, 
if this is calculated using low-energy $e^+ e^-$
annihilation data. On the other hand, the discrepancy with the Standard Model is greatly
reduced if one uses $\tau$ decay data to estimate the Standard Model contribution to
$g_\mu - 2$. Normally, one would prefer to use $e^+ e^-$ data, since they are related more
directly to $g_\mu - 2$, with no need to worry about isospin violations, etc. Until very
recently, the $e^+ e^-$ data seemed to be agreeing very nicely, but preliminary
measurements by BABAR using the radiative-return method~\cite{Davier}
do not agree so well. Until these
discrepancies get ironed out, one should take $g_\mu - 2$ {\it cum grano salis}!

\section{The Role of the Cold Dark Matter Constraint}
\label{CDM}

The most precise constraint on supersymmetry may be that provided by the density
of cold dark matter, as determined from astrophysical and cosmological measurements
by WMAP {\it et al}~\cite{WMAP}:
\begin{equation}
\Omega_{CDM} \; = \; 0.1099 \pm 0.0062 .
\label{OCDM}
\end{equation}
Applied straightforwardly to the relic LSP density $\Omega_{LSP} h^2$,
this would give a very tight relation between supersymmetric
model parameters, fixing some combination of them at the \% level. 
This would essentially reduce the dimensionality of the supersymmetric 
parameter space by one unit. Let us
assume for now that the LSP is the lightest neutralino $\chi$, whose density is
usually thought to be fixed by freeze-out from thermal equilibrium in the
early Universe, and concentrate our
attention on the minimal supersymmetric extension of the Standard Model (MSSM).
In this case, respecting the constraint (\ref{OCDM}) would force one
into narrow WMAP `strips' in planar projections of the MSSM parameters~\cite{EOSS}. However,
caution should be exercised before jumping to this conclusion.

The $\chi$ density would depend on the expansion rate at the freeze-out temperature,
which is typically about 1/25 of $m_\chi$, and hence in the multi-GeV range. It is
usually assumed that the standard Big-Bang expansion rate applied at that epoch,
but we have no evidence that this was the case before the epoch of Big-Bang
nucleosynthesis. Indeed, a successful calculation of the relic $\chi$ density using
information from the LHC and other accelerators would be the best confirmation
that standard Big-Bang cosmology applied earlier than that epoch~\cite{MN}.

\begin{figure}
\resizebox{0.45\textwidth}{!}{%
  \includegraphics{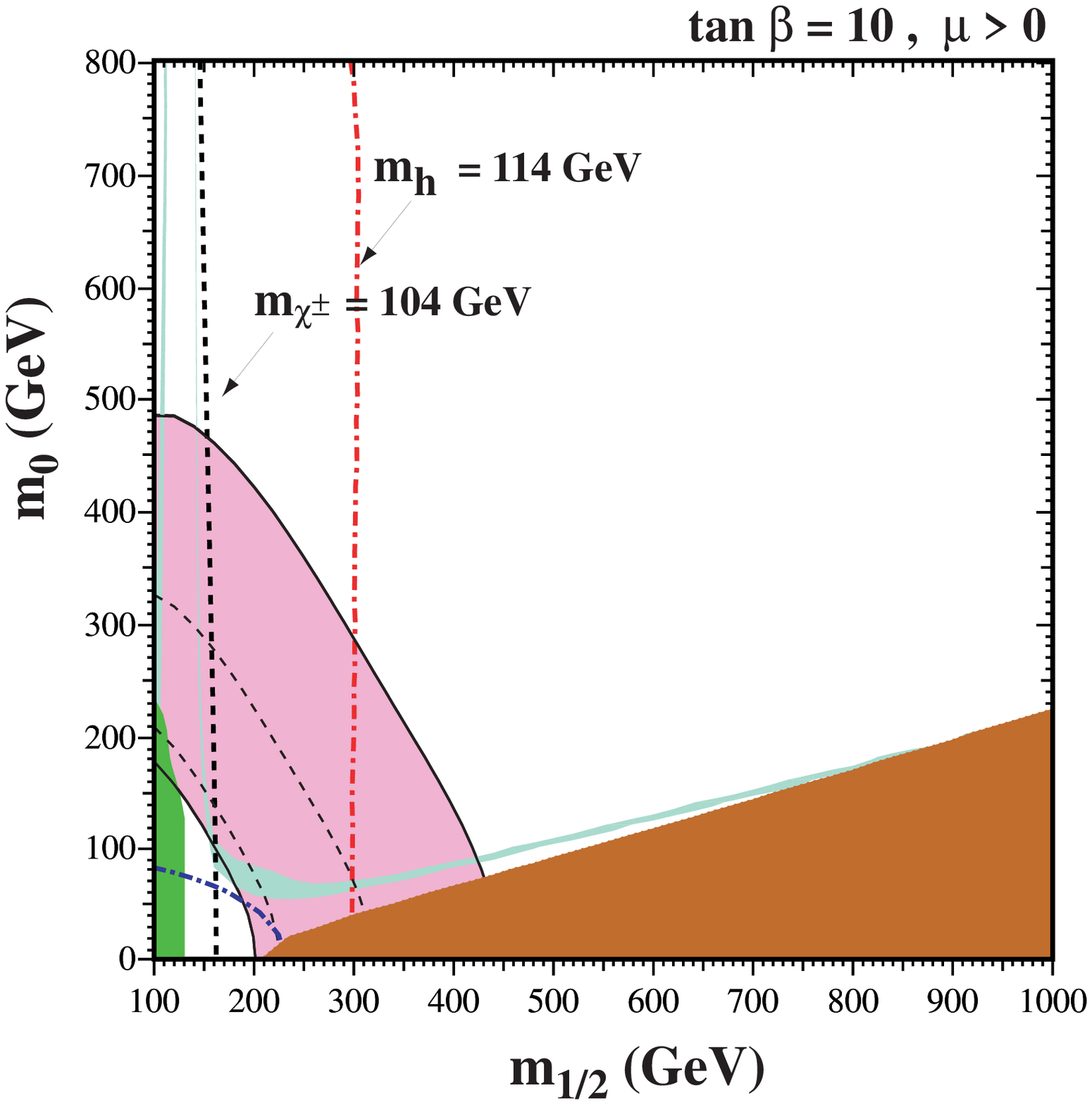}
}
\resizebox{0.45\textwidth}{!}{%
  \includegraphics{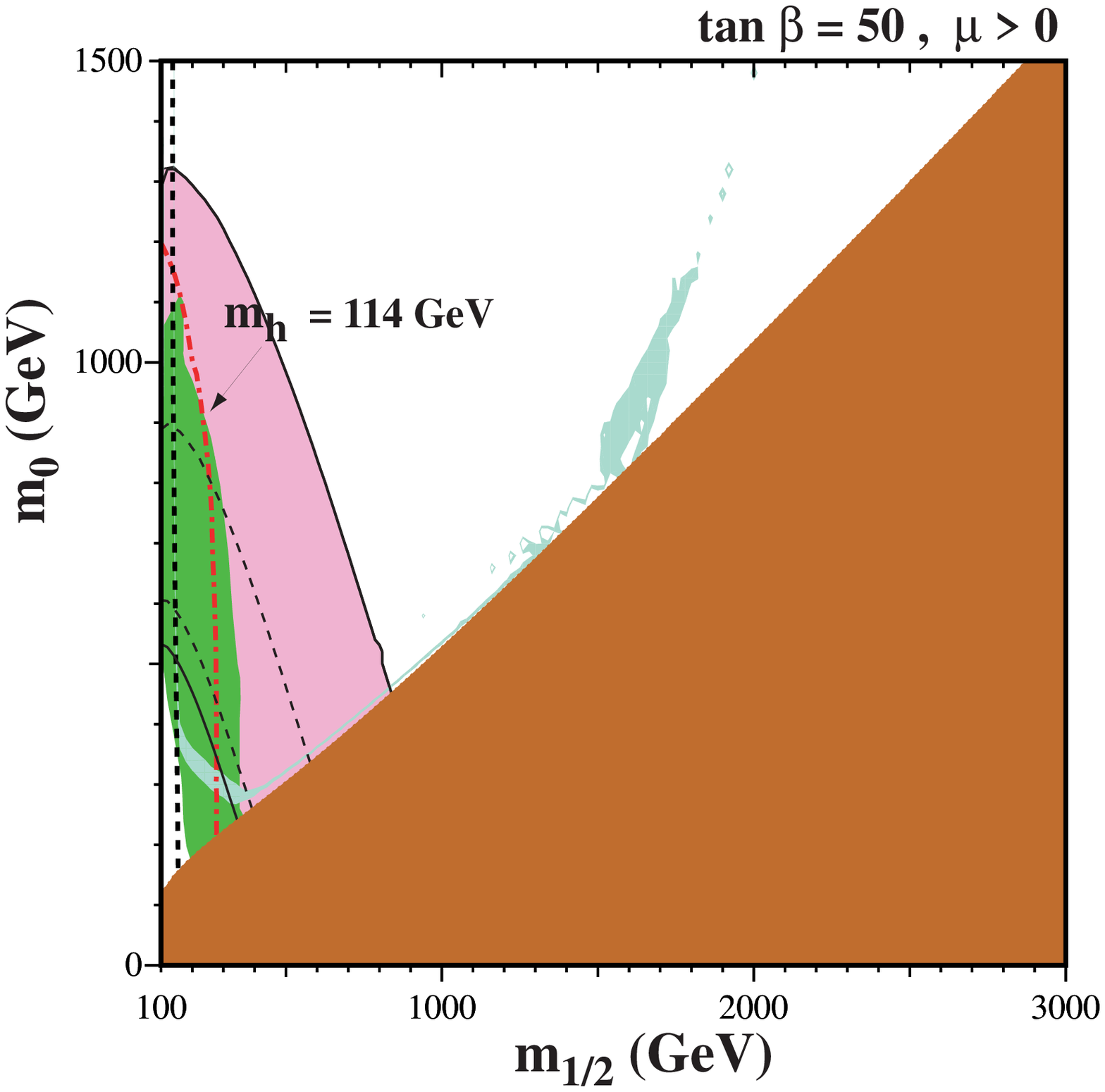}
}
\caption{The $(m_{1/2}, m_0)$ planes in the CMSSM for (a) $\tan \beta = 10$ and (b)
$\tan \beta = 50$, $\mu > 0$  and $A_0 = 0$. The brown regions are excluded because the LSP is
the lighter stau. The green regions are excluded by $b \to s \gamma$, and regions to the left
of the $m_{\chi^\pm}$ and $m_h$ contours are excluded by LEP. The regions favoured by
the cold dark matter constraint are light (turquoise), and the regions favoured by $g_\mu - 2$
are pink~\protect\cite{Olive}.}
\label{fig:CDM}       
\end{figure}

Moreover, supersymmetry might not be the only contribution to the cold dark matter,
in which case (\ref{OCDM}) should be interpreted as an upper limit on $\Omega_{LSP} h^2$.
However, most of the supersymmetric parameter
space in simple models gives a supersymmetric relic density that exceeds the WMAP range
(\ref{OCDM}), e.g., above the WMAP `strip' in Fig.~\ref{fig:CDM}(a),
and the regions with lower density generally correspond to (if anything)
{\it lower} values of the sparticle masses, e.g., below
the WMAP `strip' in Fig.~\ref{fig:CDM}(a). So this may not be a big deal.

%

However, even if one takes them seriously, the locations of these WMAP `strips' do vary 
significantly with the choices of other supersymmetric parameters, as can be seen by
comparing the cases of $\tan \beta = 10, 50$ in Fig.~\ref{fig:CDM}(a, b)~\cite{Olive}. These two plots
are in the CMSSM, where the super- symmetry-breaking scalar masses $m_0$ and
gaugino masses $m_{1/2}$ are each assumed to be universal at the GUT scale.
As one varies $\tan \beta$, the WMAP `strips' cover much of the $(m_{1/2}, m_0)$ plane.

Several different regions of the CMSSM $(m_{1/2}, m_0)$ plane can be distinguished, in
which different dynamical processes are dominant. At low values of $m_{1/2}$
and $m_0$, simple $\chi - \chi$ annihilations via crossed-channel sfermion
exchange are dominant, but this `bulk' region is now largely excluded by the LEP
lower limit on the Higgs mass, $m_h$. At larger $m_{1/2}$, but relatively small $m_0$,
close to the boundary of the region where the lighter stau is lighter than the
lightest neutralino: $m_{\tilde \tau_1} < m_\chi$, coannihilation between the $\chi$
and sleptons are important in suppressing the relic $\chi$ density into the WMAP
range (\ref{OCDM}), as seen in Fig.~\ref{fig:CDM}(a). 
At larger $m_{1/2}, m_0$ and $\tan \beta$, the relic $\chi$
density may be reduced by rapid annihilation through direct-channel $H, A$ Higgs 
bosons, as seen in Fig.~\ref{fig:CDM}(b). Finally, the relic density can again be brought
down into the WMAP range (\ref{OCDM}) at large $m_0$ not shown in 
Fig.~\ref{fig:CDM}, in the `focus-point' region close the boundary where electroweak 
symmetry breaking ceases to be possible and the lightest neutralino $\chi$
acquires a significant higgsino component.

\begin{figure}
\resizebox{0.45\textwidth}{!}{%
  \includegraphics{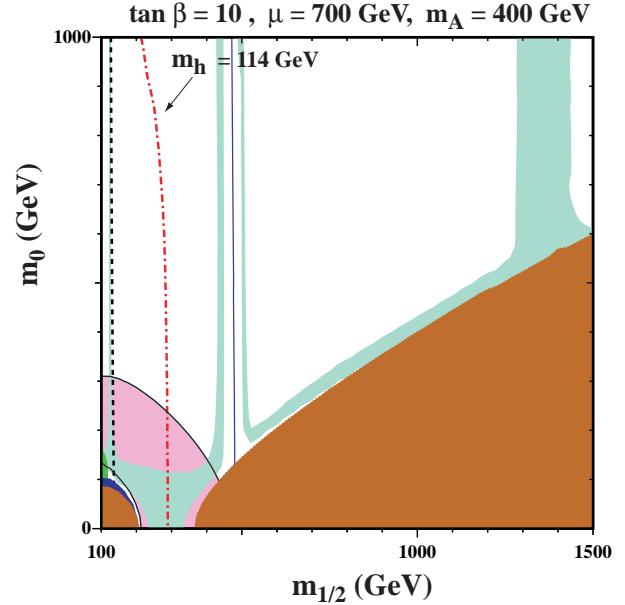}
}
\caption{The $(m_{1/2}, m_0)$ plane in the NUHM for $\tan \beta = 10$, 
$\mu = 700$~GeV and $m_A = 400$~GeV~\protect\cite{NUHM}. The colours of the shadings
and contours are the same as in Fig.~\protect\ref{fig:CDM}.}
\label{fig:NUHM}       
\end{figure}

These regions move around if one abandons the universality assumptions of the
CMSSM. For example, if one allows the supersymmetry-breaking contributions
to the Higgs mases to be non-universal (NUHM), the rapid-annihil- ation WMAP 
`strip' can appear at different values of $\tan \beta$ and $m_{1/2}$, as seen in
Fig.~\ref{fig:NUHM}~\cite{NUHM}. Rapid annihilation through the direct-channel $H, A$
poles suppresses the relic density between the two parallel vertical WMAP strips,
and the relic density is suppressed in the right-hand strip because the neutralino
LSP has a significant higgsino component.

%

\begin{figure}
\resizebox{0.45\textwidth}{!}{%
  \includegraphics{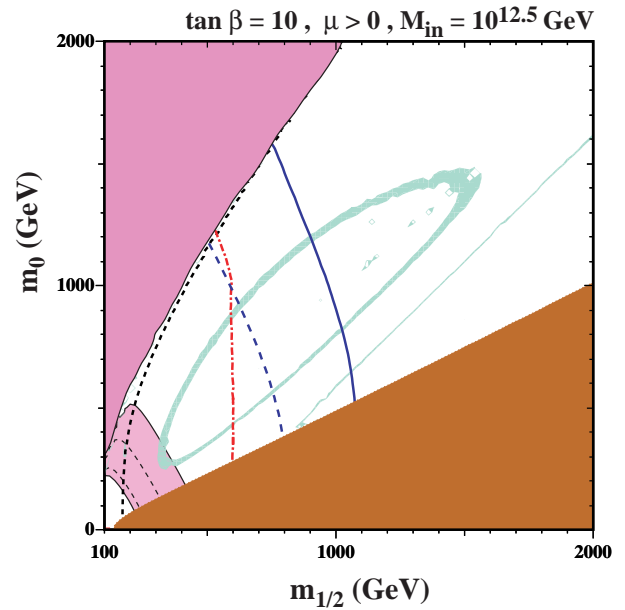}
}
\caption{The $(m_{1/2}, m_0)$ plane in a GUTless model with $\tan \beta = 10$,
$\mu > 0$  and $A_0 = 0$, assuming universality at $M_{in} = 
10^{12.5}$~GeV~\protect\cite{Pearl}. The colours of the shadings
and contours are the same as in Fig.~\protect\ref{fig:CDM}.}
\label{fig:GUTless}       
\end{figure}

The appearance of the $(m_{1/2}, m_0)$ plane is also changed significantly if one
assumes that the universality of soft super- symmetry-breaking masses in the CMSSM
occurs not at the GUT scale, but at some lower renormalization scale~\cite{Pearl}, as occurs in
some `mirage unification' models. In this case, the sparticle masses are generally
closer together, and the bulk, coannihilation,
rapid-annihilation and focus-point regions approach each other
and eventually merge as the mirage unification scale is reduced, as illustrated in
Fig.~\ref{fig:GUTless}. In such `GUTless' models, $\Omega_{LSP} h^2$ falls below
the WMAP range (\ref{OCDM}) in larger regions of the $m_{1/2}, m_0$ plane.

%

\begin{figure}
\resizebox{0.45\textwidth}{!}{%
  \includegraphics{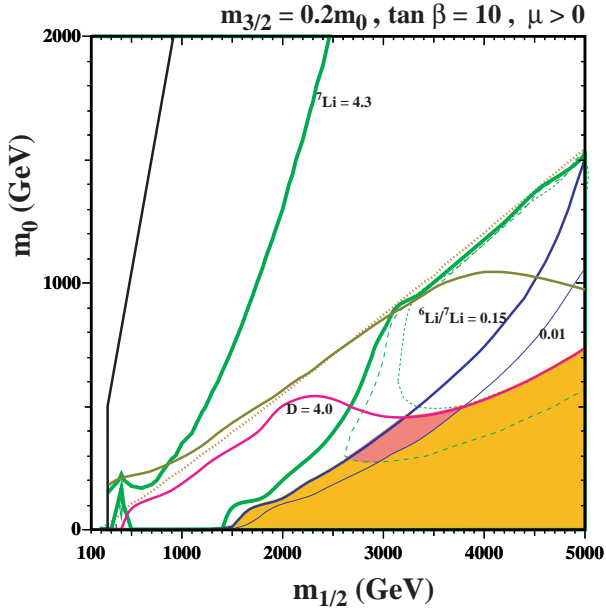}
}
\caption{The $(m_{1/2}, m_0)$ plane in a GDM model with a $\tan \beta = 10$,
$\mu > 0$  and $m_{3/2} = 0.2 m_0$~\protect\cite{CEFOS}. The contours are labelled with the
corresponding values of the light-element abundances. The pink region is
favoured by the $^7$Li abundance~\protect\cite{7Li}, and the yellow region is
consistent with the other observed light-element abundances.
}
\label{fig:GDM}       
\end{figure}

So far, it has been assumed that the lightest neutralino $\chi$ is the LSP. A sneutrino
LSP is ruled out by a combination of LEP searches that exclude lower masses and 
direct dark matter searches that exclude higher sneutrino masses. How about a
gravitino LSP (GDM)? In this case, the next-to-lightest supersymmetric particle (NLSP) must be
metastable, since it decays by gravitational-strength interactions. What might the NLSP
be in such a gravitino LSP scenario? In the CMSSM at low $m_0$ and large $m_{1/2}$
it would be the $\tilde \tau_1$, which would have a distinctive experimental
signature at the LHC, as discussed later. This GDM scenario is tightly constrained by the
astrophysical constraints on the cosmological abundances of light elements, as
seen in Fig.~\ref{fig:GDM}~\cite{CEFOS}. However, such a scenario might have some advantages,
e.g., by enabling the cosmological prediction for the abundance of $^7$Li~\cite{7Li} to be
improved, as also shown in Fig.~\ref{fig:GDM}.

%

\begin{figure}
\resizebox{0.45\textwidth}{!}{%
  \includegraphics{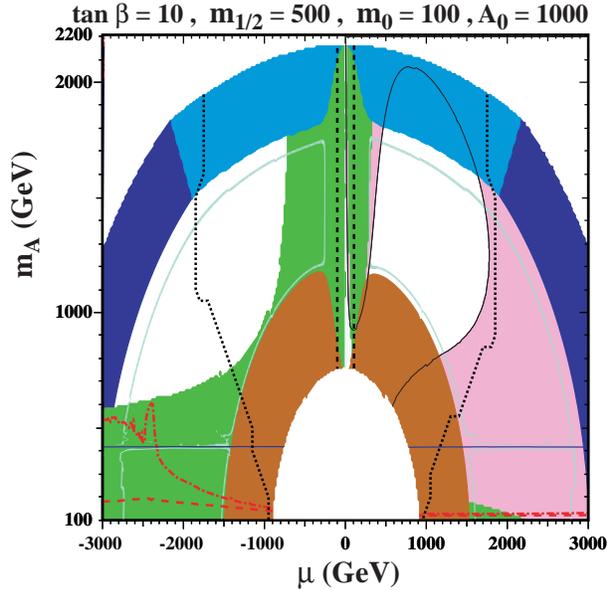}
}
\caption{The $(\mu, m_A)$ plane in the NUHM with $\tan \beta = 10$,
$m_{1/2} = 500$~GeV, $m_0 = 100$~GeV and $A_0 = 1000$~GeV~\protect\cite{sneutrino}. The
$\tau$ ($e, \mu$) sneutrino is the lightest Standard Model partner in the light
(dark) blue region, in which case cosmology requires the gravitino to be the
LSP and the sneutrino is the NLSP. The colours of the other shadings
and contours are the same as in Fig.~\protect\ref{fig:CDM}.}
\label{fig:sneutrino}       
\end{figure}

If one goes beyond the CMSSM, there are other possibilities for the NLSP in a 
gravitino LSP scenario. For example, within the NUHM the NLSP might be the 
lighter stop squark, ${\tilde t_1}$~\cite{stop}, which would also have a spectacular signature 
at the LHC, as also discussed later. Alternatively, the NLSP could be one of the
sneutrinos~\cite{sneutrino}, as illustrated in Fig.~\ref{fig:sneutrino}.

%

\begin{figure}
\resizebox{0.45\textwidth}{!}{%
  \includegraphics{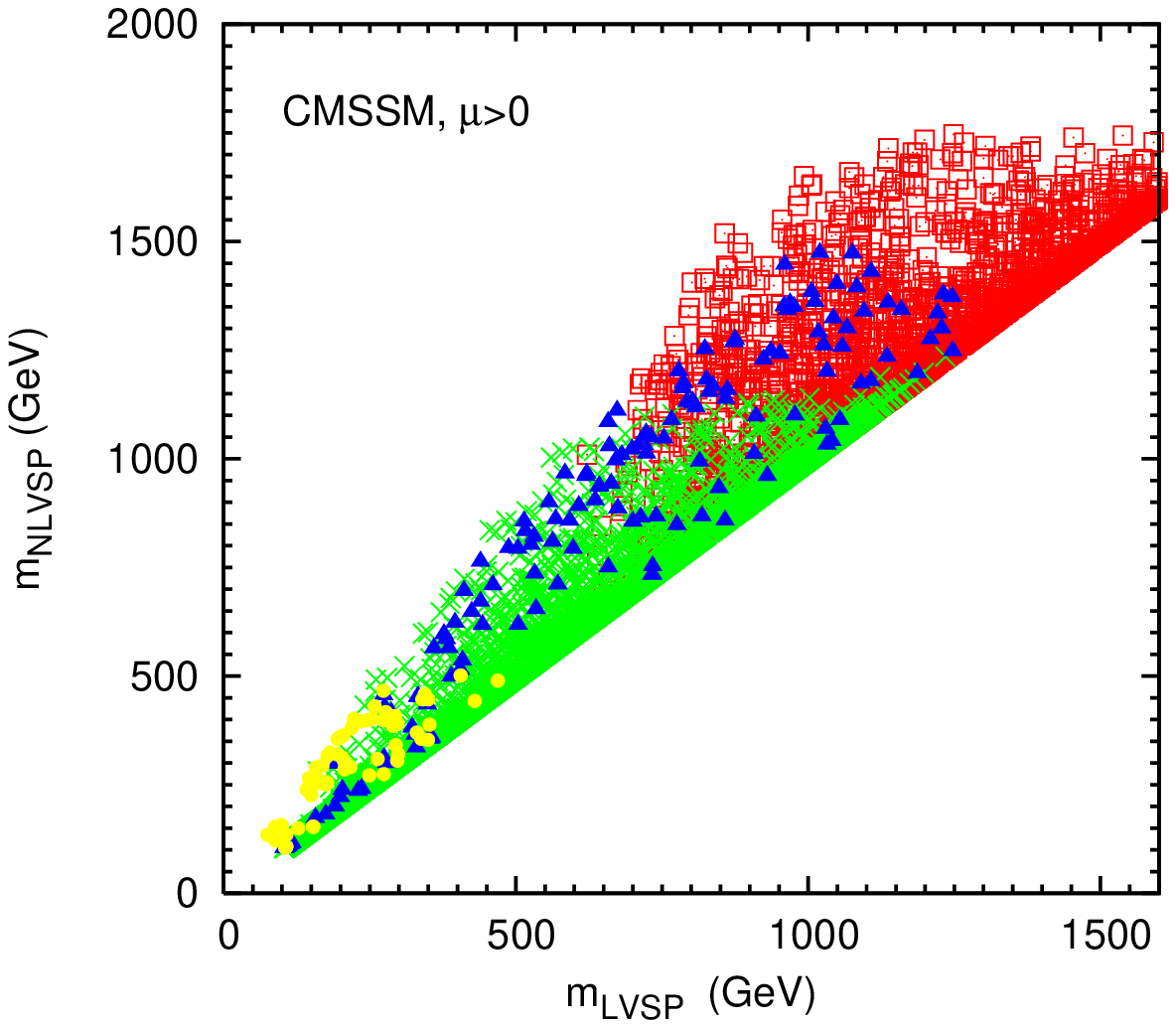}
}
\resizebox{0.45\textwidth}{!}{%
  \includegraphics{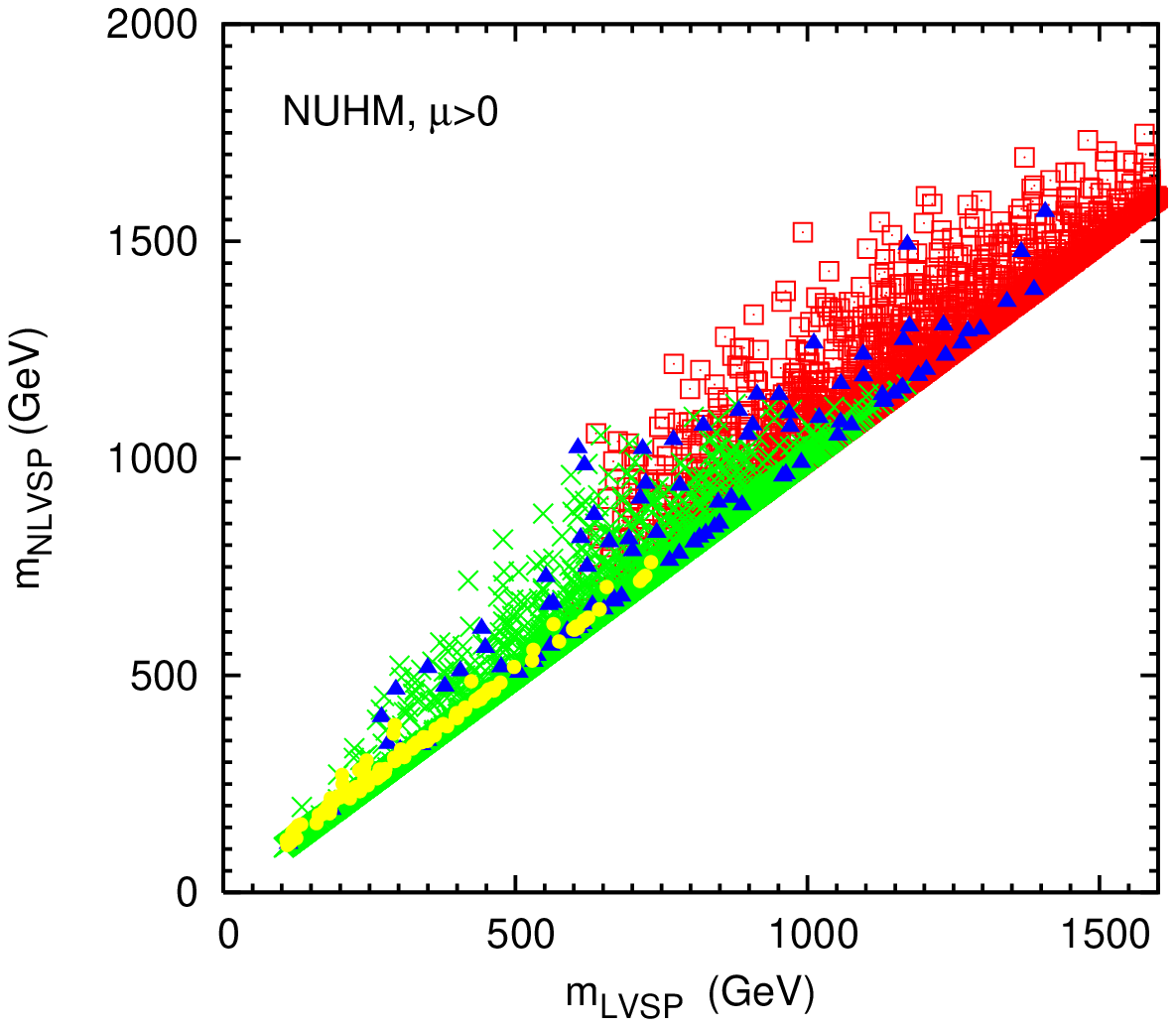}
}
\caption{Scatter plots of the lightest and next-to-lightest visible supersymmetric
particles in (a) the CMSSM and (b) the NUHM~\protect\cite{LVSP}. The colours are explained in
the text. Most (but not all) of the dark matter points are accessible to the LHC.}
\label{fig:scatter}       
\end{figure}

As this brief review has indicated, there are many ways in which supersymmetry
may choose to obey the WMAP constraint (even assuming that $R$ parity is
conserved). Considering specifically the CMSSM and the NUHM, and
assuming that the LSP is the lightest neutralino $\chi$, what are the
prospects for discovering supersymmetry at the LHC in light of WMAP? In the
two panels of Fig.~\ref{fig:scatter} you see large samples of CMSSM and NUHM
models, respectively~\cite{LVSP}. The full sample is shown in red, the blue points are
compatible with WMAP, the green points are accessible to the LHC, and the yellow
points are accessible to direct dark matter searches. It is apparent that the LHC
will be able to explore most, but not all, of the WMAP-compatible models, and its
prospects look brighter than those for direct dark matter searches. However, there
is no guarantee that supersymmetry will be within the reach of the LHC.

%

\section{Results from a Global Likelihood Analysis}

If one wishes to go further in assessing the prospects for discovering
supersymmetry at the LHC, one must introduce some supplementary
considerations. A theoretical argument that is often invoked is that of
minimizing the amount of fine-tuning required to obtain the electroweak
scale and/or the right cold dark matter density. It is true that, generally
speaking the heavier the sparticles, the more fine-tuning is required~\cite{FT}.
Although this argument is encouraging for the LHC, how much 
fine-tuning is too much? It is difficult to state an objective criterion.
Alternatively, let us not feed in any theoretical prejudice, but instead
ask the available experimental data, and make likelihood 
analyses of the parameter spaces of simple
supersymmetric models, such as the CMSSM or the NUHM.

In addition to WMAP {\it et al.}, the available data include
electroweak precision observables (EWPO), B-decay observables, and
$g_\mu - 2$. The EWPO give no firm evidence for any new physics beyond
the Standard Model: $m_W$ is somewhat higher than might have
been expected, but there is no indication from $\sin^2 \theta$ or other
EWPO. Likewise, the B-decay observables, including $b \to s \gamma$,
$B \to \tau \nu$, $B_d$ and $B_s$ mixing and $B_s \to \mu^+ \mu^-$
decay give only upper limits on the possible effects of supersymmetry.

As already mentioned,
the only real experimental hint for new physics comes from $g_\mu - 2$. In the analysis
of this Section we assume the following discrepancy between experiment~\cite{g-2}
and the Standard Model calculation based on $e^+ e^-$ data~\cite{g-2th}:
\begin{equation}
a_\mu^{exp} - a_\mu^{SM} \; = \; (30.2 \pm 8.8) \times 10^{-10},
\label{g-2}
\end{equation}
with an additional theoretical uncertainty of $2 \times 10^{-10}$. I
emphasize again that this number has not yet settled down~\cite{Davier}, and
specifically that the discrepancy would be much reduced if one
used $\tau$ data. Nevertheless, what are the prospect for the LHC if
one incorporates (\ref{g-2}) into a global likelihood analysis?

We have recently published such an analysis~\cite{Master}, using a Markov-Chain
Monte Carlo (MCMC) technique to explore efficiently the likelihood function in
the parameter spaces of the CMSSM and the NUHM1 (the simplest version
of the NUHM, in which the two Higgs soft supersymmetry-breaking masses
are assumed equal)~\cite{Bayes}. A full list of the observables and the values assumed
for them in this global analysis are given in~\cite{Master1}, as updated 
in~\cite{Master}. 

\begin{figure}
\resizebox{0.45\textwidth}{!}{%
  \includegraphics{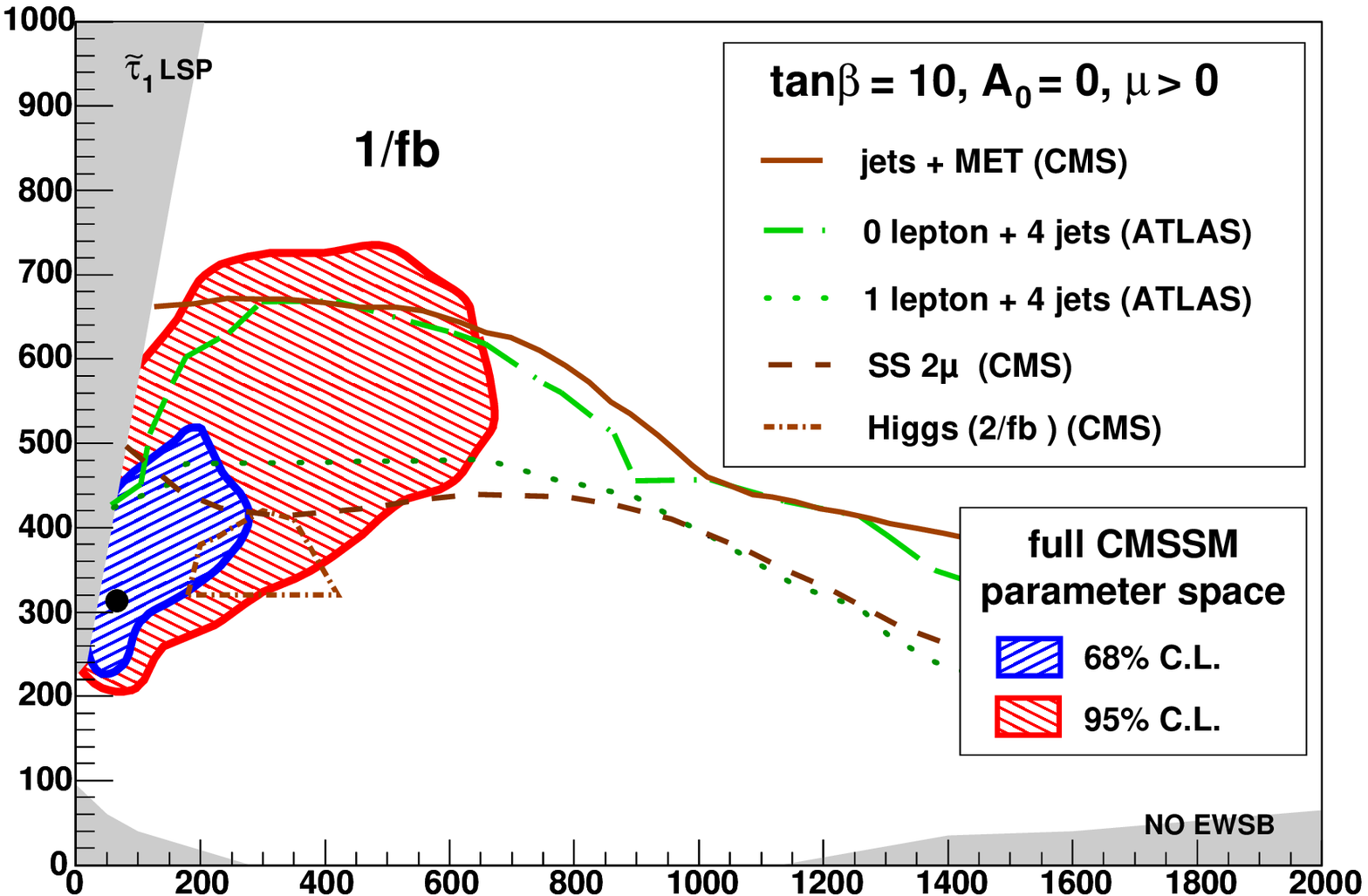}
}
\resizebox{0.45\textwidth}{!}{%
  \includegraphics{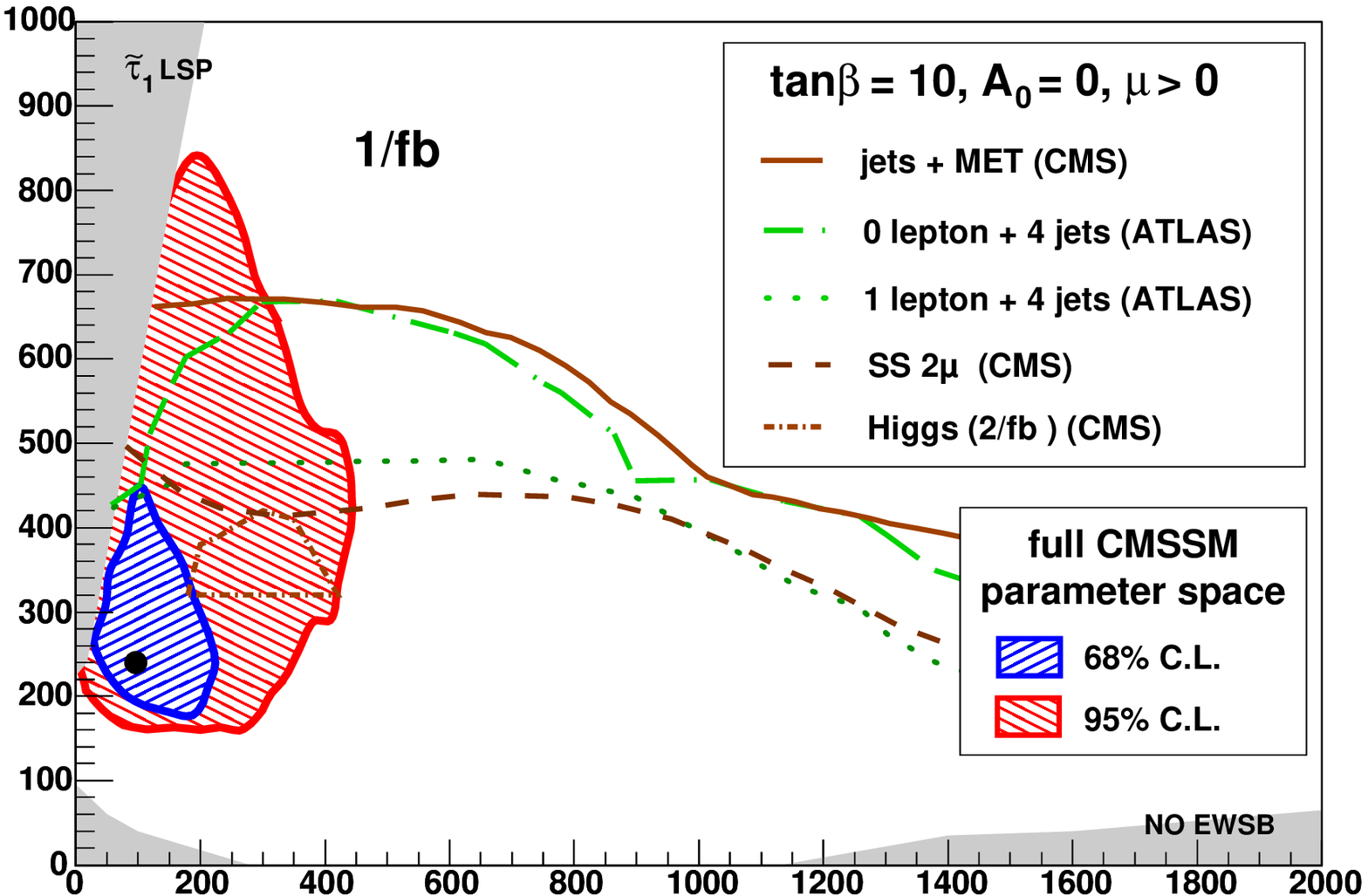}
}
\caption{The $(m_0, m_{1/2})$ planes in (a) the CMSSM and (b)
the NUHM1, showing the regions favoured in a likelihood analysis
at the 68\% (blue) and 95\% (red) confidence levels~\protect\cite{Master}. The best-fit
points are shown as black points. Also shown are
the discovery contours in different channels for the LHC with 1/fb
(2/fb for the Higgs search in cascade decays of sparticles)~\protect\cite{LHC}.}
\label{fig:MCMC}       
\end{figure}

The 68\% and 95\% confidence-level (C.L.) regions in the
$(m_{1/2}, m_0)$ planes of the CMSSM and NUHM1 are shown in
Fig.~\ref{fig:MCMC}~\cite{Master}. Also shown for comparison are the physics
reaches of ATLAS and CMS with 1/fb of integrated luminosity~\cite{LHC}. 
(MET stands for missing
transverse energy, SS stands for same-sign dilepton pairs, and the
sensitivity for finding the lightest Higgs boson in cascade decays of
supersymmetric particles is calculated for 2/fb of data.) Our likelihood
analysis assumes $\mu > 0$, as motivated by the sign of the
apparent discrepancy in $g_\mu - 2$, but sampled all values of $\tan \beta$
and $A_0$: the experimental sensitivities were estimated
assuming $\tan \beta = 10$ and $A_0 = 0$, but are probably not
very sensitive to these assumptions. The global maxima of the
likelihood function (indicated by the black dots) are as follows: 
in the CMSSM $m_{1/2} = 310$~GeV,
$m_0 = 60$~GeV, $A_0 = 240$~GeV, $\tan \beta = 11$ and
$\chi^2/N_{dof} = 20.4/19$ (37\% probability), and in the
NUHM1 $m_{1/2} = 240$~GeV, $m_0 = 100$~GeV, 
$A_0 = - 930$~GeV, $\tan \beta = 7$ and $\chi^2/N_{dof} = 18.0$ 
(39\% probability). It is encouraging that the best-fit points lie
well within the LHC discovery range, as do the 68\% and most of the
95\% C.L. regions. It is also encouraging that the two best-fit points
have similar values of $m_{1/2}, m_0$ and $\tan \beta$, the most
important parameters for the sparticle spectrum, indicating that the 
likelihood analysis is relatively insensitive to the theoretical model
assumptions.

%

\begin{figure}
\resizebox{0.45\textwidth}{!}{%
  \includegraphics{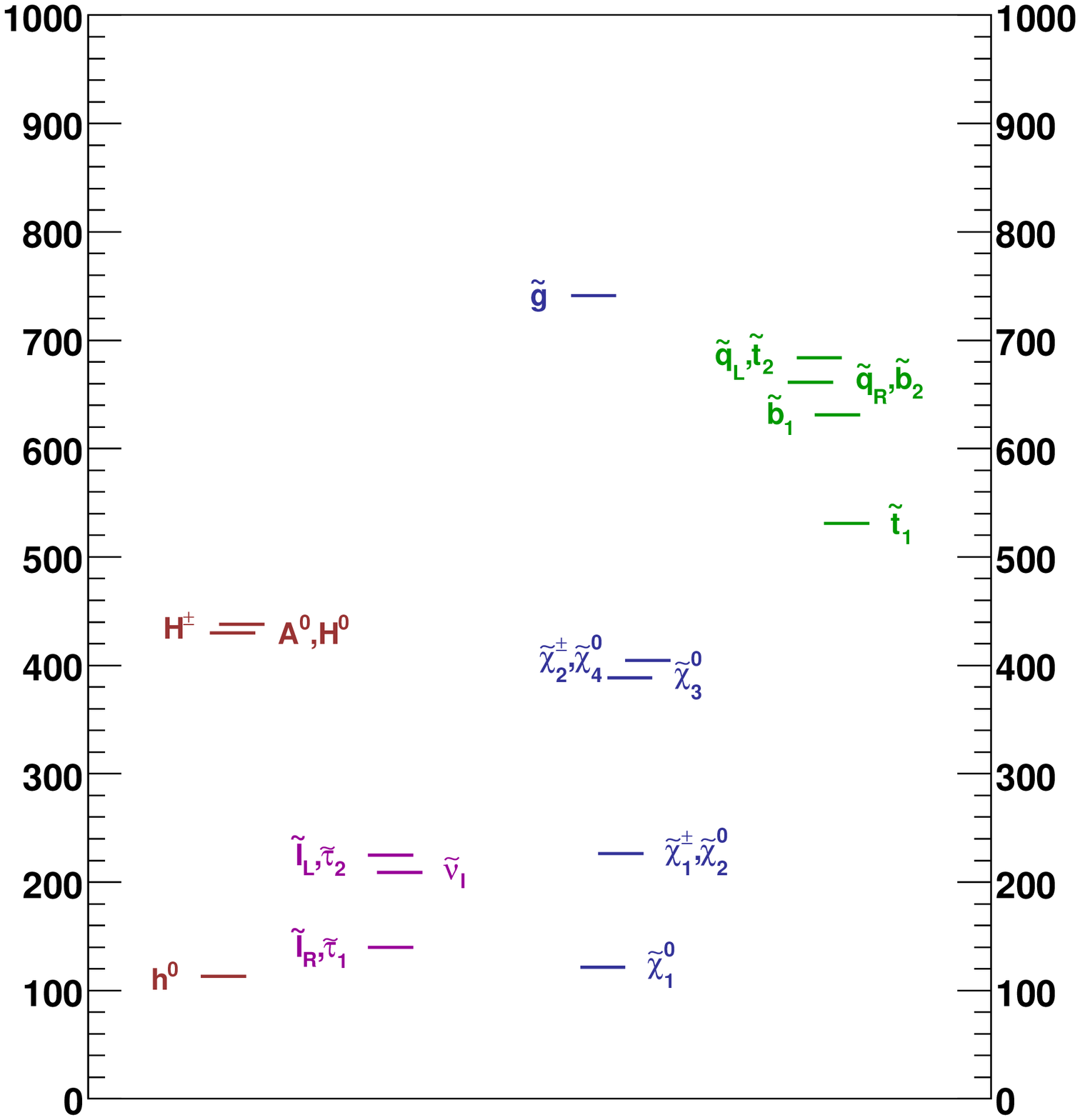}
}
\resizebox{0.45\textwidth}{!}{%
  \includegraphics{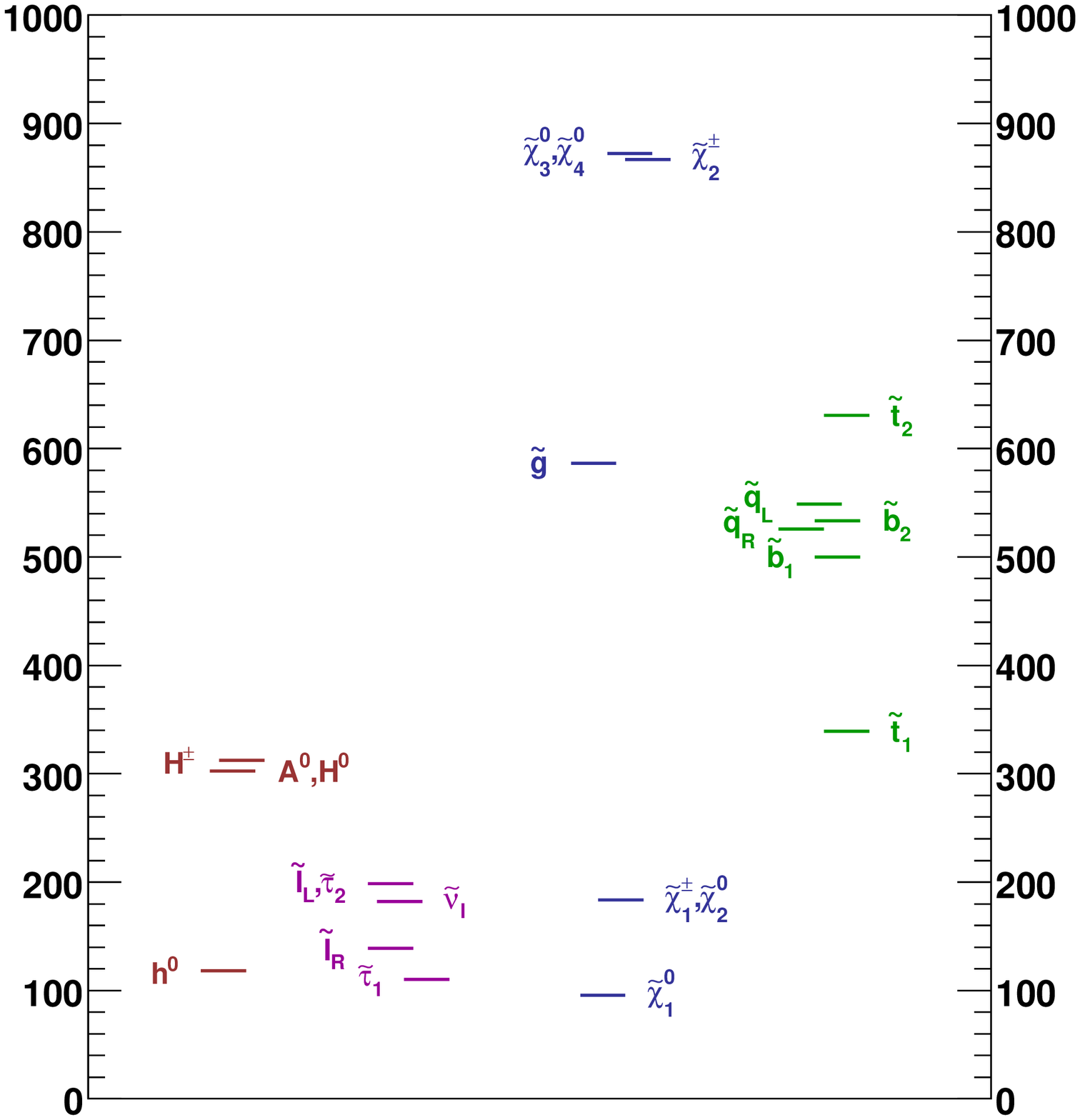}
}
\caption{Spectra at the best-fit points in Fig.~\protect\ref{fig:MCMC}(a) and (b),
for the CMSSM and NUHM1, respectively~\protect\cite{Master}.}
\label{fig:spectra}       
\end{figure}

The sparticle spectra for the best fits in the CMSSM and NUHM1
are displayed in Fig.~\ref{fig:spectra}~\cite{Master}. The main differences are the
following. The lightest Higgs boson is appreciably lighter in the CMSSM,
113~GeV vs 118~GeV (indeed this is the main reason the $\chi^2$ of
the CMSSM fit is higher, and it is acceptable only because of the
theoretical uncertainty in the calculation of $m_h$), whereas the
heavier MSSM Higgs bosons are lighter in the NUHM1. The heavier
neutralinos and charginos are significantly heavier in the NUHM1 than
in the CMSSM, reflecting the larger best-fit value of $\mu$ (870~GeV
compared to 380~GeV). The gluino and squarks are somewhat
lighter in the NUHM1, reflecting the smaller value of $m_{1/2}$.

%

\begin{figure}
\resizebox{0.45\textwidth}{!}{%
  \includegraphics{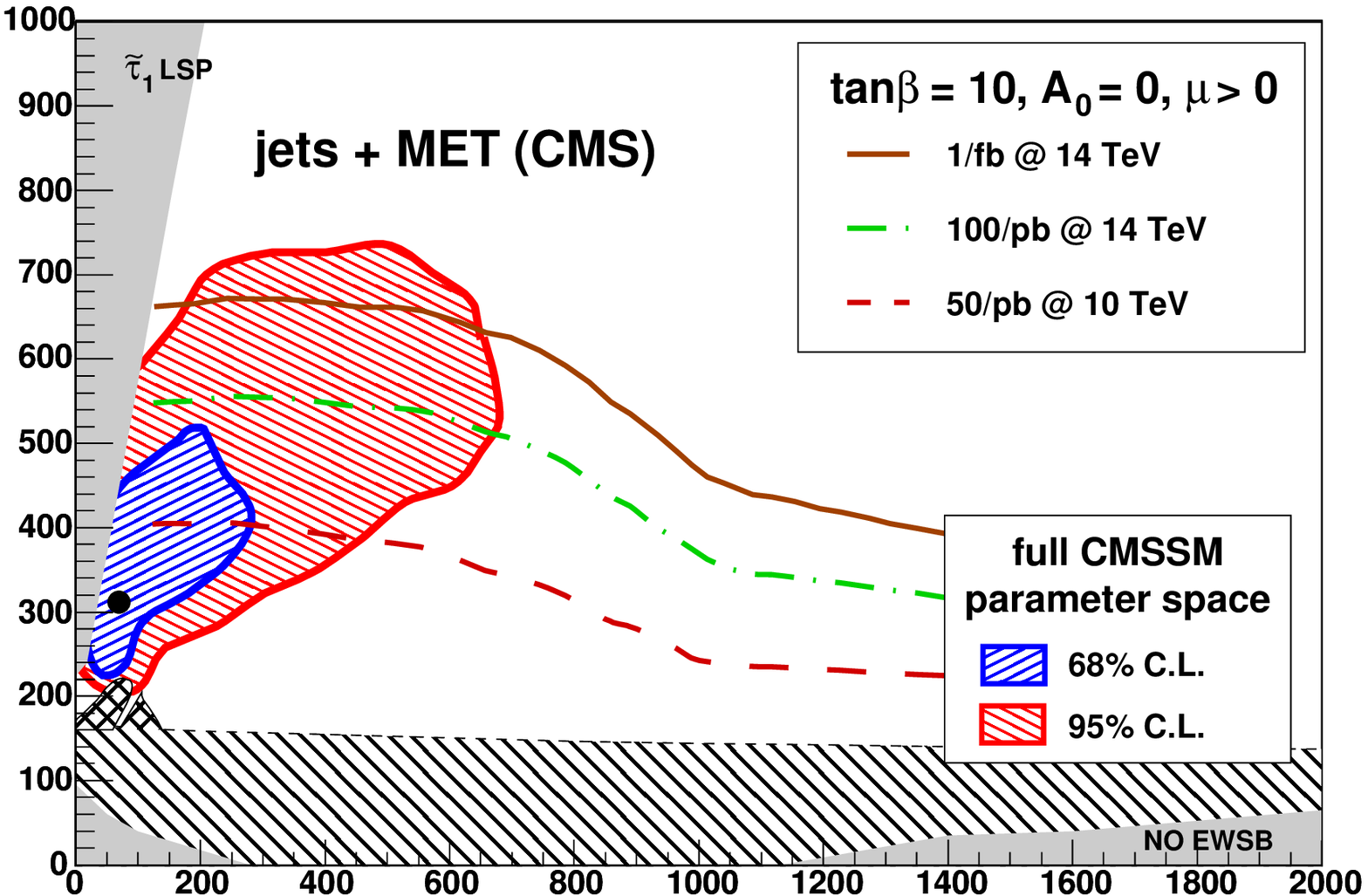}
}
\resizebox{0.45\textwidth}{!}{%
  \includegraphics{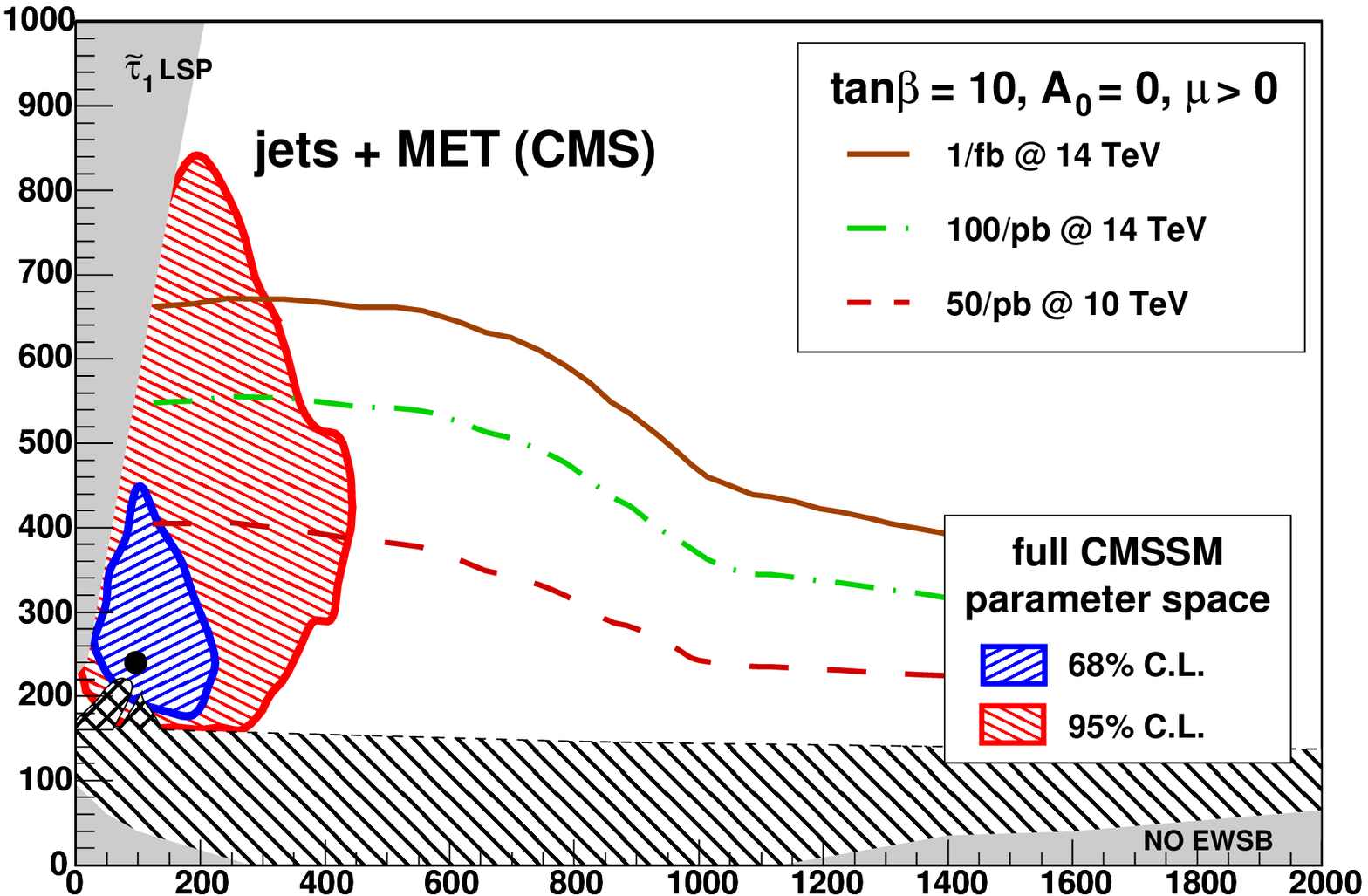}
}
\caption{Results of the likelihood analysis shown in Fig.~\protect\ref{fig:MCMC}~\protect\cite{Master},
now displaying also the LHC discovery contours for 1/fb at 14~TeV,
100/pb at 14~TeV, and 50/pb at 10~TeV.}
\label{fig:lumi}       
\end{figure}

According to this analysis,
the prospects for discovering supersymmetry at the LHC look quite
bright, in both the CMSSM and the NUHM1. How soon might the discovery be
made? As seen in Fig.~\ref{fig:lumi}, the best-fit points in both the
CMSSM and the NUHM1 lie within the region accessible to the LHC
with 50/pb of integrated luminosity at 10~TeV, and the 68\% C.L. 
regions lie within the reach of the LHC with 100/pb of data. (The
hatched regions at the bottom of each panel of Fig.~\ref{fig:lumi} are
those excluded by searches for supersymmetry at LEP and the
Tevatron collider.) We see that even modest LHC luminosity, not
necessarily at the design energy, would already explore substantial
new regions of the CMSSM and NUHM1 parameter spaces, and
might already have a good chance of discovering supersymmetry!

%

\begin{figure}
\resizebox{0.45\textwidth}{!}{%
  \includegraphics{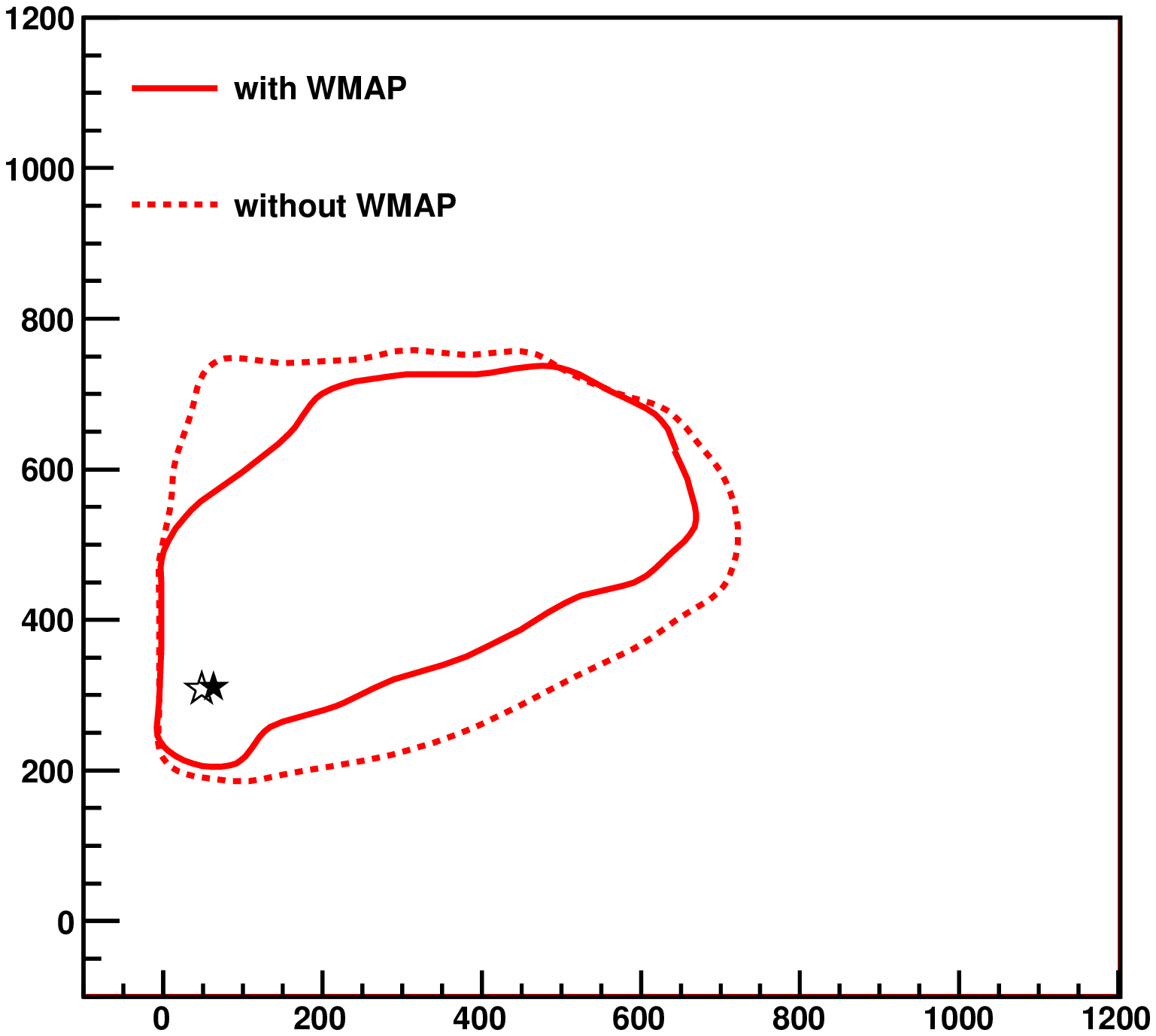}
}
\resizebox{0.45\textwidth}{!}{%
  \includegraphics{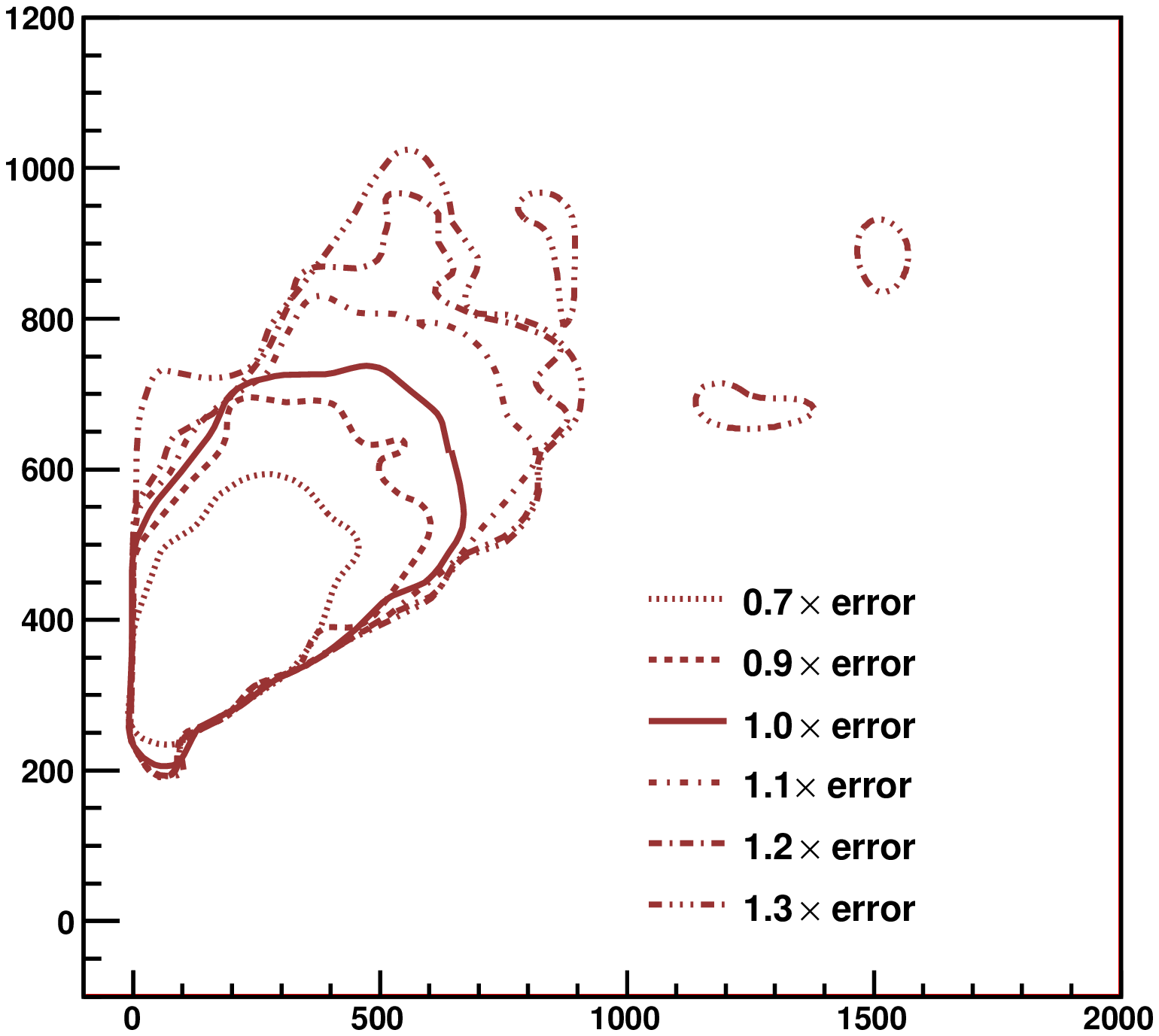}
}
\caption{The $(m_0, m_{1/2})$ plane in the CMSSM likelihood analysis,
showing the effects (a) of removing the WMAP constraint, and (b) of
varying the estimated error in the $g_\mu - 2$ constraint compared to the
default value~\protect\cite{Master}.}
\label{fig:WMAP}       
\end{figure}

How dependent are these results on the inputs used? 
Fig.~\ref{fig:WMAP}(a) shows the effect on the 95\% C.L. region in the
CMSSM $(m_{1/2}, m_0)$ plane of removing the WMAP constraint~\cite{Master}.
There is surprisingly little effect, basically because
(although they are individually very narrow) the WMAP
strips for different values of $\tan \beta$ and $A_0$ largely cover the
$(m_{1/2}, m_0)$ plane. Hence, an analysis such as ours that
samples different values of $\tan \beta$ and $A_0$ does not spot
the reduction in the dimensionality of the parameter space. On the
other hand, we see in Fig.~\ref{fig:WMAP}(b) that relaxing the $g_\mu - 2$
constraint even slightly, by increasing the relative error, has a
dramatic effect on the preferred region of the CMSSM 
$(m_{1/2}, m_0)$ plane~\cite{Master}. If the $g_\mu - 2$ constraint was relaxed
much more, it would encompass the focus-point region, which is
disfavoured in the default version of our MCMC likelihood analysis.

%

\section{Alternative Supersymmetric Phenomenologies}

All the above discussion has assumed that $R$ parity is
conserved, in which case the LSP is stable and hence must be 
neutral, leading to a missing-energy signature at the LHC. If $R$
parity is not conserved, the dark matter constraint evaporates,
and sparticle decays may be observable at the LHC. For
example, the neutralino may decay visibly into a combination
of three jets and/or leptons.

\begin{figure}
\resizebox{0.45\textwidth}{!}{%
  \includegraphics{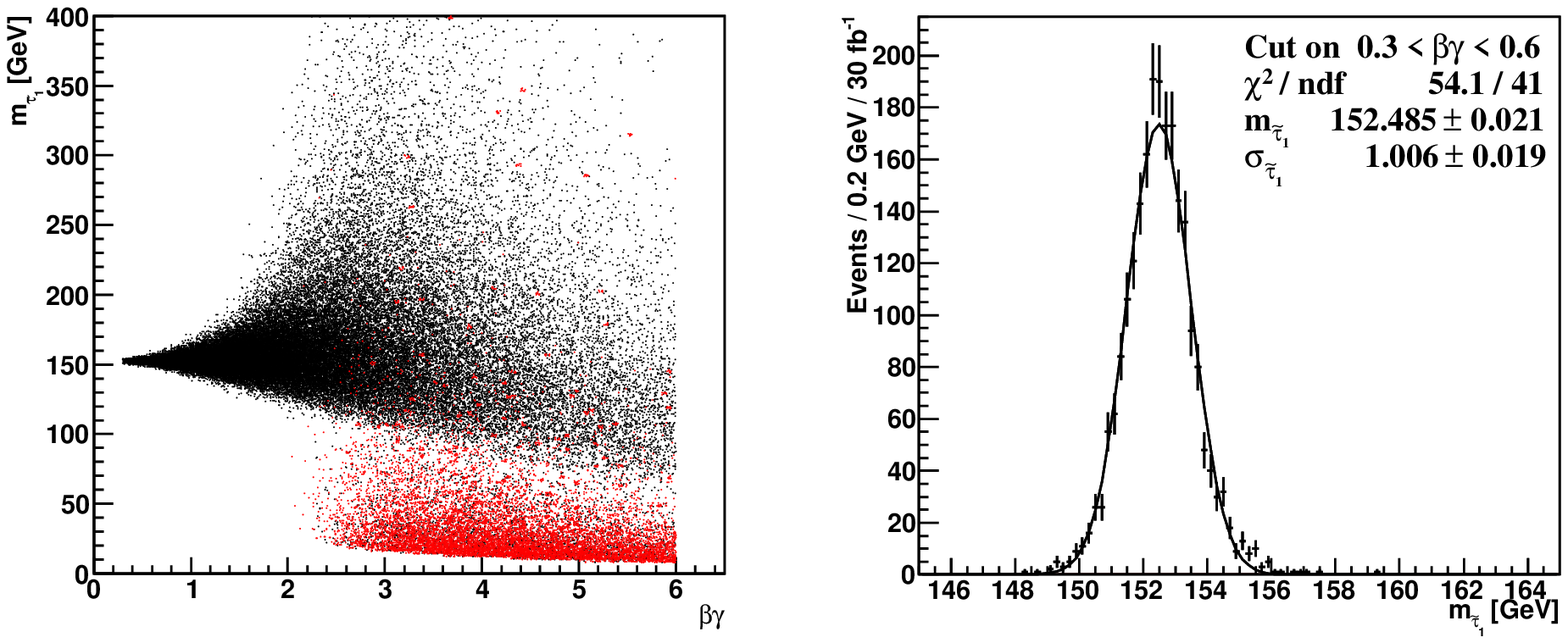}
}
\resizebox{0.45\textwidth}{!}{%
  \includegraphics{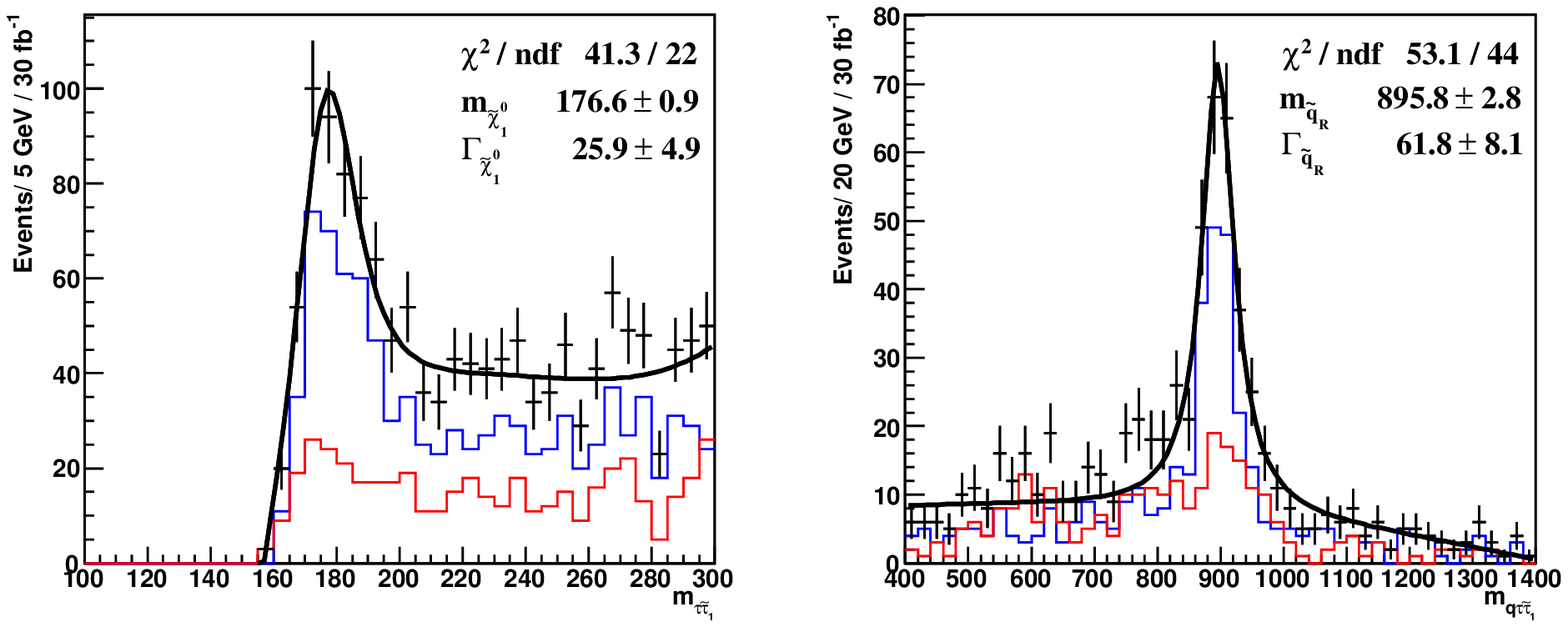}
}
\caption{Results of an analysis~\protect\cite{ERO} showing that the ATLAS
detector could (upper panels) measure accurately the stau NLSP
mass, and (lower panels) reconstruct the $\chi$ and $\tilde q_R$ masses
in cascade decays in a GDM scenario.} 
\label{fig:stau}       
\end{figure}

A less radical possibility is that $R$ parity is conserved, but
the LSP is the gravitino, in which case the NLSP is metastable.
The NLSP might be the lighter stau ${\tilde \tau_1}$, which
would be easily distinguishable as a non-relativistic charged
particle. Studies have shown that the mass of the ${\tilde \tau_1}$
could be measured very accurately, and that one could easily
reconstruct heavier sparticles that decay into the ${\tilde \tau_1}$,
as seen in Fig.~\ref{fig:stau}~\cite{ERO}.

%

Alternatively, the NLSP might be the lighter stop ${\tilde t_1}$~\cite{stop}.
Immediately after production at the LHC, it would become confined
inside a charged or neutral hadron. As it moves through an LHC
detector, it would have a high probability of changing its charge
as it interacts with the material in the detector. This combined with
its non-relativistic velocity would provide a truly distinctive
signature.

Yet another possibility is that the NLSP might be some flavour of
sneutrino~\cite{sneutrino}, in which case the characteristic signature would be
missing energy carried away by the metastable sneutrino. This
could nevertheless be distinguished from the conventional
missing-energy signature of a neutralino LSP (or NLSP),
because the final states would be more likely to include the
charged lepton with the same flavour as the sneutrino NLSP, 
either $e$, $\mu$ or $\tau$.

These are just a few examples of the possible alternatives to
the conventional missing-energy signature of supersymmetry.
Studies have shown that the LHC would also have good prospects
for detecting such signatures.

\section{Playing for High Stakes}

The start-up of the LHC resembles a high-stakes game of roulette.
We have good reasons for expecting the LHC will reveal the
mechanism of electroweak symmetry breaking, whether it is the
Higgs boson of the Standard Model or something more complicated.
Supersymmetry is just one of the theories jostling for consideration as
physics beyond the Standard Model that might be revealed by the LHC.

As I discussed earlier, I believe there are many good motivations
for expecting supersymmetry to appear at the TeV scale. On the other
hand, I think it was Feynman who once remarked that if you had one 
good reason you would not need to give any more! Anyway, I believe
that there are excellent prospects for producing supersymmetry at the
LHC, and equally excellent prospects for detecting it if it is produced.

The stakes in the LHC supersymmetry search are certainly high: it
would be a completely novel symmetry of Nature, linking bosons and
fermions; it could be considered as circumstantial evidence for string
theory; it could pave the way to unification of the fundamental forces;
it could stabilize the puzzling hierarchy of mass scales in fundamental
physics; it might explain 80\% of the matter in the Universe. LHC
roulette is not a game for the faint-hearted: let the protons turn!

\section*{Acknowledgements}

I gratefully acknowledge my many collaborators on the topics discussed here,
particularly Keith Olive.

\end{document}